\documentclass[11pt]{article}
\usepackage[dvips]{graphicx}  
\begin{document}

\title{\LARGE \bf  PROCESS PHYSICS: INERTIA, 
 GRAVITY\\ and the QUANTUM\footnote{ Contribution to the 3rd Australasian
Conference on
 General Relativity and Gravitation, Perth, Australia, July 2001}  }  
\author{{Reginald T. Cahill}\\
  {School of Chemistry, Physics and Earth Sciences}\\{ Flinders University
}\\ { GPO Box
2100, Adelaide 5001, Australia }\\{(Reg.Cahill@flinders.edu.au)}}

\date{}
\maketitle

\begin{abstract}
{\it  Process Physics} models reality as self-organising 
relational or semantic information using a self-referentially limited neural
network model. This generalises the traditional non-process  syntactical
modelling of reality by taking account of  the limitations  and
characteristics of self-referential syntactical information  systems,
discovered by  G\"{o}del  and  Chaitin, and the analogies with the standard
quantum formalism and its limitations.   In process physics space and quantum physics
are emergent  and unified, and time is a distinct non-geometric process.  Quantum
phenomena are caused by fractal topological   defects  embedded in and
forming a growing three-dimensional fractal process-space.   Various 
features of the emergent physics are briefly discussed including:   quantum
gravity, quantum field theory, limited causality
and  the Born quantum measurement metarule,  inertia,
time-dilation effects, gravity and the equivalence principle, a growing universe with  a cosmological
constant,  black holes and event horizons, and the emergence of classicality.
\end{abstract}

\vspace{20mm}
\hspace{5mm} Key words:  process physics, G\"{o}del's theorem, neural network, semantic 

\hspace{27mm}information, self-referential noise, process-time, process-space,

\hspace{27mm}quantum gravity. 

\newpage

\section{Introduction}
There is mounting evidence that a unification of gravity and quantum theory   has finally
been achieved, but only after the realisation that the failure to do so, until recently, arose
from deep limitations to the traditional modelling of reality by physicists.  From its
earliest inception physics has modelled reality using  formal or syntactical
information systems. These have undefined {\it a priori} entities, such a fields and
geometry, together with   {\it a priori} laws.  However the actual structure has
always been a little messier than that: it  is in all cases composed of mathematical
equations  supplying the core structure, together with metarules that make up for
deficiencies of the mathematical  model, and finally some metaphysical assertions that
usually have an ontological flavour.

 A simple early example is actually Newton's geometrical modelling of the
phenomena of time. There the mathematical structure is the one-dimensional
continuum or geometrical  line, but that fails because it has no matching for the present
moment effect or even the distinction between past and future. The geometrical-time metarule
here involves the notion that one must actively imagine a point moving along a line at a
uniform rate, and in this regard the metarule is certainly not inconsistent with the
mathematical model. Finally the metaphysical assertion is that time {\it is} geometrical.
Clearly then from the beginning physicists have blurred the three components of modelling. 

The ongoing failure of physics to fully match all the aspects
of the phenomena of time, apart from that of order,  arises because physics has always
used non-process models, as is the nature of formal or syntactical systems. Such  systems
do not require any notion of process - they are entirely structural and static. The new
process physics
\cite{CK97,CK98,CK99,CKK00} overcomes these deficiencies my using a non-geometric
process model for time (see \cite{MC} for an early  non-technical account), but process physics  also  argues
for the importance of  relational or semantic information in modelling reality. Semantic information
refers to the notion that reality is a purely informational system where the information is
internally meaningful: to be more specific such information has the form of
self-organising patterns which also generate their own `rules of interaction'. In this way we
see the correctness of Wheeler's insight of `Law without Law'\cite{Wheeler}.  Hence the
information is `content addressable', rather than is the case in the usual syntactical
information modelling where the information is represented by symbols.  This symbolic or
syntactical mode is only applicable to higher level phenomenological descriptions, and has
served physics well. 

 A pure semantic information system must be  formed by
a subtle bootstrap process. The mathematical model for this has the form of a stochastic
neural network (SNN)  for the simple reason that neural networks are well known for their
pattern or non-symbolic information  processing abilities\cite{neural}.  The stochastic
behaviour is related to the limitations of syntactical systems discovered by
G\"{o}del\cite{G} and more recently extended  by
Chaitin\cite{Chaitin90,Chaitin99,Chaitin01}, but also results in the neural network being
innovative in that it creates its own patterns.  The neural network  is self-referential,
and   the stochastic input,  known as self-referential noise, acts both to limit the depth
of the self-referencing   and also to generate potential order. 

Herein is a status report on the ongoing development of process physics beginning with a
 discussion of the comparison of syntactical and the new semantic information system
and their connections with G\"{o}del's incompleteness theorem. Later sections describe
the emergent unification of gravitational and quantum phenomena, amounting to a quantum
theory of gravity.

\section{Syntactical and Semantic Information Systems}

In modelling reality with  formal or syntactical information systems  physicists assume that
the full behaviour of a physical system can be compressed into axioms and rules for the
manipulation of symbols. However G\"{o}del discovered that self-referential syntactical
systems (and these includes basic mathematics) have fundamental
limitations which amount to the realisation that not all truths  can be compressed into an
axiomatic structure, that formal systems are much weaker than previously supposed.  In physics
such systems have always been used in conjunction with metarules and metaphysical assertions, all
being `outside' the formal system and designed to overcome the limitations of the syntax. Fig.1
depicts the current understanding of self-referential syntactical systems. Here the key feature is
the G\"{o}del boundary demarcating the provable from the unprovable truths of some system. 
Chaitin has demonstrated that in mathematics the unprovable truths are essentially random in
character.  This, however, is a structural randomness in the sense that the individual truths do
not have any structure to them which could be exploited to  condense them down to or be encoded in
axioms. This is unlike random physical events which occur in time.  Of course syntactical
systems are based on the syntax of symbols and this is essentially non-process or
non-timelike.

\vspace{-10mm}

\begin{figure}[h]
\hspace{17mm}
\setlength{\unitlength}{1.75mm}
\hspace{20mm}\begin{picture}(40,25)
\thicklines
\qbezier(0,10)(20,30)(40,10)
\qbezier(0,0)(20,-20)(40,0)
\qbezier(0,0)(-4,5)(0,10)
\qbezier(40,10)(44,5)(40,0)

\put(19.7,-10){\line(0,1){30}}

\qbezier(0,0)(15,5)(0,10)

\put(-1,6){{\bf Formal}}
\put(-1,3){{\bf  Axioms}}

\qbezier[60](5,2)(15,-1)(24,0)
\put(24,0){\vector(4,1){1}}
\put(16,-1){\line(1,2){1.5}}
\put(5.0,-3.0){\bf G\"{o}del's Thm}
\put(25.5,0.2){\circle*{0.5}}
\put(20,-2.2){\bf unprovable truth}

\put(13,14){\circle*{0.5}}

\put(7,15){\bf  provable truth}
\qbezier(5,8)(7.5,12.5)(13,14)
\put(12,13.71){\vector(2,1){0.1}}

\put(22.6,5){\circle*{0.5}}
\put(22.6,8){\circle*{0.5}}
\put(27.6,10){\circle*{0.5}}
\put(25.6,3.5){\circle*{0.5}}
\put(24,6){\bf random  truths}
\put(26,4.0){\bf (Chaitin)}

\put(11,-7){\bf THE BOUNDARY}

\end{picture}
\vspace{15mm}\caption{\small Graphical depiction of the `logic space' of
a self-referential syntactical information  system, showing the formal system
consisting of symbols and rules, and an example of one theorem
 (a provable truth).  Also shown  are unprovable truths which in
general are random (or unstructured) in character, following the work of Chaitin.
The G\"{o}delian boundary is the demarcation between provable
and unprovable truths. 
 \label{section:figure:Godel}}
\end{figure}
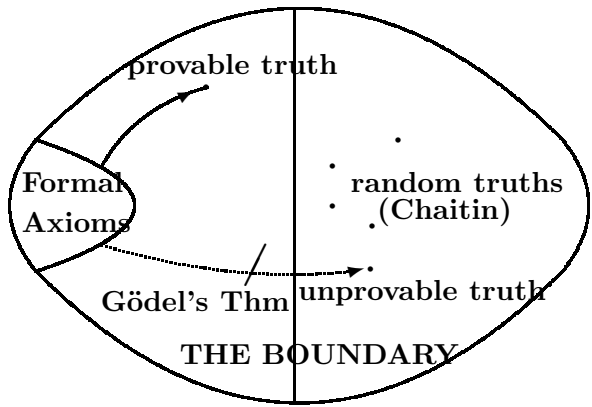

There is an analogy between the structure of self-referential syntactical
information systems and the present structure of quantum theory as depicted in Fig.2.
There the formal and hence non-process mathematical structure is capable of producing
many provable truths, such as the energy levels of the hydrogen atom, and these are also true in
the sense that they agree with reality.  But from the beginning of quantum theory the Born
measurement metarule was introduced to relate this non-process modelling to the actual randomness
of quantum measurement events.  The individuality of such random events is not a part of the
formal structure of quantum theory. Of course it is well known that  the non-process or 
structural aspects of the probability metarule  are consistent with the mathematical formalism, in
the form of the usual `conservation of probability' equation and the like.  Further, the  quantum
theory has always  been subject to various metaphysical interpretations, although these
have never played a key role for practitioners of the theory.  This all suggests that
perhaps the Born metarule is bridging a G\"{o}del-type boundary, that there is a bigger
system required to fully model quantum aspects of reality, and that the boundary is evidence of
self-referencing in that system.

\vspace{-10mm}

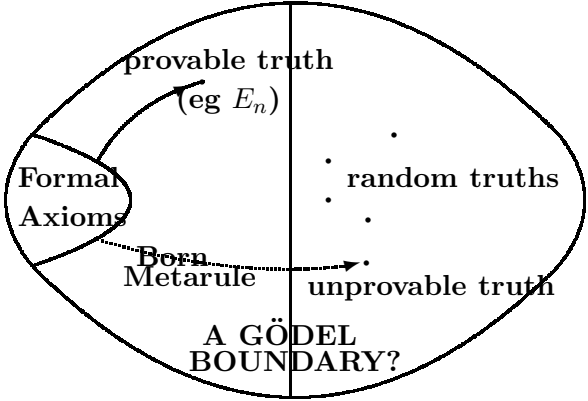
\begin{figure}[h]
\hspace{17mm}
\setlength{\unitlength}{1.75mm}
\hspace{20mm}\begin{picture}(40,25)
\thicklines
\qbezier(0,10)(20,30)(40,10)
\qbezier(0,0)(20,-20)(40,0)
\qbezier(0,0)(-4,5)(0,10)
\qbezier(40,10)(44,5)(40,0)

\put(19.7,-10){\line(0,1){30}}

\qbezier(0,0)(15,5)(0,10)

\put(-1,6){{\bf Formal}}
\put(-1,3){{\bf  Axioms}}

\qbezier[60](5,2)(15,-1)(24,0)
\put(24,0){\vector(4,1){1}}
\put(8.0,0.0){\bf Born}
\put(7.0,-1.5){\bf Metarule}
\put(25.5,0.2){\circle*{0.5}}
\put(21,-2.2){\bf unprovable truth}

\put(13,14){\circle*{0.5}}

\put(7,15){\bf  provable truth}
\put(11,12){\bf (eg $E_n$)}
\qbezier(5,8)(7.5,12.5)(13,14)
\put(12,13.71){\vector(2,1){0.1}}

\put(22.6,5){\circle*{0.5}}
\put(22.6,8){\circle*{0.5}}
\put(27.6,10){\circle*{0.5}}
\put(25.6,3.5){\circle*{0.5}}
\put(24,6){\bf random  truths}

\put(13,-6){\bf A G\"{O}DEL }\put(12,-8){\bf BOUNDARY? }

\end{picture}
\vspace{15mm}\caption{\small Graphical depiction of the syntactical form of
conventional quantum theory.  The Born measurement metarule appears to bridge a
G\"{o}del-like boundary.
\label{section:figure:Quantum}}

\end{figure}

Together the successes and failures of 
physics suggest that  a generalisation of  the traditional use of  syntactical 
information theory is required to model reality, and that this has now been identified as a
semantic information system which has the form of a stochastic neural network.

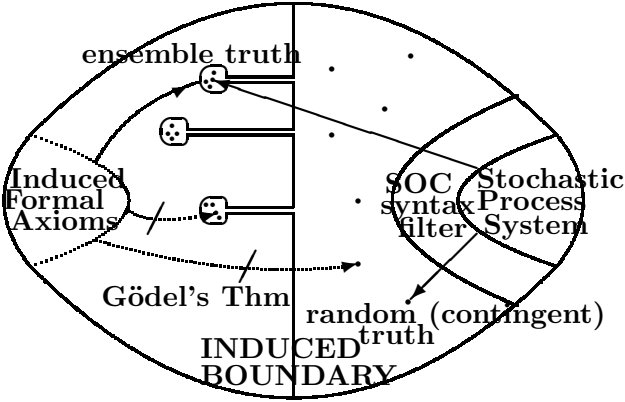
\begin{figure}[h]
\vspace{-10mm}
\setlength{\unitlength}{1.75mm}

\hspace{40mm}\begin{picture}(40,25)
\thicklines
\qbezier(0,10)(20,30)(40,10)
\qbezier(0,0)(20,-20)(40,0)
\qbezier(0,0)(-4,5)(0,10)
\qbezier(40,10)(44,5)(40,0)


\put(20.1,-10){\line(0,1){14}}
\put(20.1,4.45){\line(0,1){5.6}}
\put(20.1,10.4){\line(0,1){3.6}}
\put(20.1,14.4){\line(0,1){5.6}}

\qbezier[70](0,0)(15,5)(0,10)
\qbezier(40,0)(25,5)(40,10)
\qbezier(37,-3)(18,5)(37,13)

\put(-1.5,6.0){{\bf Induced}}
\put(-2.0,4.35){{\bf  Formal}}
\put(-1.5,2.8){{\bf Axioms}}

\qbezier[60](5,2)(15,-1)(24,0)
\put(24,0){\vector(4,1){1}}
\put(16,-1.2){\line(1,2){1.2}}
\put(5.5,-3.0){\bf G\"{o}del's Thm}
\put(25,0.2){\circle*{0.5}}

\qbezier[20](7.4,4.3)(9,3)(14,4)
\put(13.28,3.8){\vector(4,1){1}}
\put(9,2.5){\line(1,2){1.2}}

\put(14,14.25){\oval(2,2)[t,l]}
\put(14,14.25){\oval(2,2)[b,l]}
\put(14,14.5){\oval(2,1.5)[t,r]}
\put(14,14.0){\oval(2,1.5)[b,r]}
\put(13.6,14.0){{\bf .}}\put(13.1,13.9){{\bf .}}
\put(13.7,14.6){{\bf .}}\put(13.4,13.5){{\bf .}}
\put(14.9,14.4){\line(1,0){5.25}}
\put(15,14){\line(1,0){5.15}}

\put(4,15.5){\bf ensemble truth}
\qbezier[70](5,8)(7.5,12.5)(12.8,14)
\put(12,13.71){\vector(2,1){0.1}}

\put(34,7.5){\vector(-3,1){20}}

\put(34,6.0){{\bf Stochastic}}
\put(34,4.25){{\bf Process}}
\put(34.5,2.5){{\bf System}}
\put(27.0,5.6){{\bf SOC}}
\put(26.7,4.02){{\bf syntax}}
\put(28,2.2){{\bf filter}}
\put(34,2.5){\vector(-1,-1){5}}\put(28.8,-2.7){\circle*{0.5}}
\put(21,-4.3){\bf random (contingent) }
\put(25,-5.9){\bf  truth}

\put(13,-7){\bf INDUCED}
\put(13,-9){\bf BOUNDARY}


\put(11,10.25){\oval(2,2)[t,l]}
\put(11,10.25){\oval(2,2)[b,l]}
\put(11,10.5){\oval(2,1.5)[t,r]}
\put(11,10.0){\oval(2,1.5)[b,r]}
\put(11.92,10.4){\line(1,0){8.25}}
\put(11.92,10){\line(1,0){8.25}}
\put(10.9,9.9){{\bf .}}\put(10.2,9.9){{\bf .}}
\put(10.5,10.5){{\bf .}}\put(10.5,9.5){{\bf .}}

\put(14,4.25){\oval(2,2)[t,l]}
\put(14,4.25){\oval(2,2)[b,l]}
\put(14,4.5){\oval(2,1.5)[t,r]}
\put(14,4.0){\oval(2,1.5)[b,r]}
\put(14.1,3.5){{\bf .}}\put(13.1,4.49){{\bf .}}
\put(13.6,4.5){{\bf .}}\put(13.85,3.9){{\bf .}}
\put(14.9,4.5){\line(1,0){5.2}}
\put(14.9,4){\line(1,0){5.25}}

\put(25,5){\circle*{0.5}}
\put(23,10){\circle*{0.5}}
\put(27,12){\circle*{0.5}}
\put(23,15){\circle*{0.5}}
\put(29,16){\circle*{0.5}}
\end{picture}
\vspace{15mm}
\caption{\small Graphical depiction of the bootstrapping of and the emergent structure of a
self-organising pure semantic information system.  As a high level effect  we see the
emergence of an induced formal system, corresponding to the current standard syntactical
modelling of reality. There is an emergent  G\"{o}del-type boundary which represents the 
inaccessibility of the random or contingent truths from the induced formal or syntactical 
system.       
\label{section:figure:Process}}
\end{figure}

Fig.3 shows a graphical depiction of the bootstrapping of a pure 
semantic information  system, showing the stochastic    neural network-like process
system from which the semantic system is seeded or bootstrapped.  Via a
Self-Organised Criticality Filter (SOCF) this seeding  system is removed or hidden.
  From the process system,  driven by
Self-Referential Noise (SRN), there are  emergent
truths, some of which are generically true (ensemble truths)  while others are 
purely contingent.  The
ensemble truths are also reachable from the
Induced Formal System as theorems, but from which, because of the non-process nature of the
induced formal system,  the contingent truths cannot be reached. In this manner there arises  a
G\"{o}del-type boundary.  The existence of the latter leads to    induced metarules 
that enhance  the
induced formal system, if that is to be used solely  in higher order phenomenology.

\section{Self-Referentially Limited  Neural Networks}
\label{sect:NN}

\begin{figure}
\setlength{\unitlength}{0.5mm}
\hspace{20mm}
\begin{picture}(150,60)
\thicklines
\put(10,10){\circle{15}}\put(8,7){1}
\put(30,30){\circle{15}}\put(27,27){2}
\put(48,0){\circle{15}}\put(45,-3){3}
\put(15,15){\vector(1,1){10}}\put(30,12){$B_{23}>0$}
\put(5,15){\vector(-1,1){10}}
\put(12,3){\vector(1,-4){3}}
\put(35,25){\vector(1,-2){9.5}}
\put(33,37){\vector(1,2){7}}
\put(37,32){\vector(3,1){15}}
\put(55,3){\vector(3,1){15}}
\put(34,-15){\vector(1,1){9}}
\put(95,15){\circle{15}}\put(94,12){i}
\put(100,20){\vector(1,1){10}}
\put(90,20){\vector(-1,1){10}}
\put(25,-22){(a)}   \put(115,-22){(b)}    \put(210,-22){(c)}

\put(145,15){\circle{15}}\put(144,12){i}
\put(150,20){\vector(1,1){10}}
\put(140,20){\vector(-1,1){10}}
\put(95,9){\oval(8,20)[b]}
\put(96,-1){\vector(1,0){1.0}} 
\put(115,12){$\in$}
\end{picture}
\hspace{40mm}
\setlength{\unitlength}{0.20mm}

\hspace{105mm}
\begin{picture}(0,50)(40,0)  
\thicklines
\put(155,165){\line(3,-5){60}}
\put(155,165){\line(-3,-5){60}}
\put(115,100){\line(3,-5){42}}
\put(195,100){\line(-3,-5){21}}
\put(135,160){ \bf $i$}
\put(225,160){ \bf $D_0\equiv 1$}
\put(225,100){ \bf $D_1=2$}
\put(225,60){ \bf $D_2=4$}
\put(225,25){ \bf $D_3=1$}
\put(155,165){\circle*{5}}
\put(115,100){\circle*{5}}
\put(195,100){\circle*{5}}
\put(95,65){\circle*{5}}
\put(135,65){\circle*{5}}
\put(175,65){\circle*{5}}
\put(215,65){\circle*{5}}
\put(155,30){\circle*{5}}
\end{picture}

\caption{\small (a) Graphical depiction of the neural network with links
$B_{ij}\in {\cal R}$ between nodes or pseudo-objects. Arrows indicate sign of
$B_{ij}$. (b) Self-links are internal to a node, so $B_{ii}=0$. (c) An $N=8$ spanning
tree for a random graph (not shown) with $L=3$.  The  distance distribution $D_k$ is indicated for
node {\it i}.
\label{section:figure:neural}}
\end{figure}
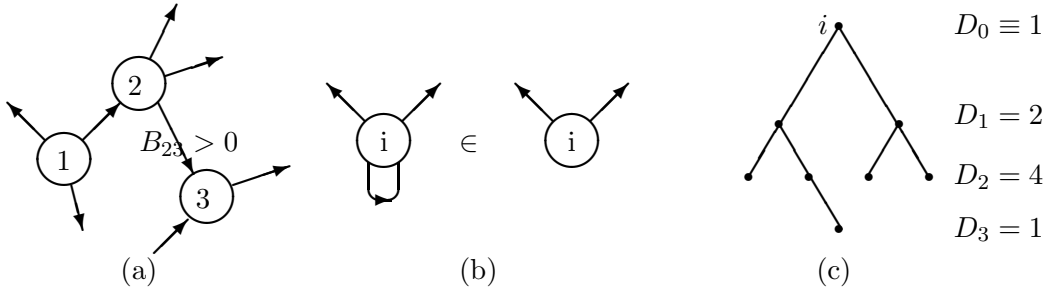

Here we briefly describe a model for a  self-referentially limited neural network and in the
following section we describe how such a network results in emergent quantum behaviour, and which,
increasingly, appears to be a unification of space and quantum phenomena. Process physics is  a semantic
information system and is devoid of {\it a priori} objects and their laws  and so it requires a subtle
bootstrap mechanism to set it up. We use a stochastic neural network, Fig.4a, having the structure of 
  real-number valued connections or relational information strengths $B_{ij}$ (considered as forming a
square matrix) between  pairs of nodes or pseudo-objects
$i$ and $j$. In standard 
neural networks\cite{neural} the network  information resides in both link and node
variables,  with the semantic information residing in attractors of the iterative network.
Such systems are also not pure in that there is an assumed underlying and manifest {\it a
priori} structure.
  
 The nodes and their link variables  will be revealed  to be themselves sub-networks of informational
relations. To avoid explicit self-connections
$B_{ii}\neq 0$, which are a part of the sub-network content of
$i$, we use antisymmetry $B_{ij}=-B_{ji}$ to conveniently ensure that 
$B_{ii}=0$, see Fig.4b.

At this stage we are using a syntactical system with symbols $B_{ij}$ and, later, rules for the
changes in the values of these variables. This system is the syntactical seed for the pure
semantic system.   Then to ensure that the nodes and links are not remnant {\it a priori}
objects the system must generate strongly linked  nodes (in the sense that the $B_{ij}$ for
these nodes are much larger than the $B_{ij}$ values for non- or weakly-linked nodes) forming
a fractal network; then self-consistently the start-up nodes and links may themselves be
considered   as mere names for sub-networks of relations.  For a successful suppression  the scheme must
display self-organised criticality (SOC)\cite{SOC} which acts as a filter for the start-up syntax. The
designation `pure' 
 refers to the notion that all seeding syntax has been removed. SOC is the process
where the emergent behaviour  displays universal criticality in that the  behaviour is
independent of the individual start-up syntax; such a start-up syntax  then  has no 
ontological significance.  

To generate a fractal structure we must  use a non-linear iterative system for the
$B_{ij}$ values.  These iterations amount to the  necessity to introduce a time-like
 process.   Any system possessing {\it a priori}  `objects' can never
be fundamental as the explanation of such objects must be outside  the system.  Hence in
process physics the absence of intrinsic undefined objects is linked with the phenomena of
time, involving as it does an ordering of `states', the present moment effect, and the
distinction between past and present. Conversely in non-process physics the presence of {\it a
priori } objects is related to the use of the non-process geometrical model of time, with this modelling
and its geometrical-time metarule being an approximate emergent description from process-time.  In this
way process physics arrives at a new modelling of time, {\it process time}, which is much more complex
than that introduced by Galileo, developed by Newton, and reaching its high point with Einstein's
spacetime geometrical model. 

The  stochastic neural network   so far has been realised with one
particular  scheme involving a stochastic non-linear matrix iteration, see (1). 
The matrix inversion $B^{-1}$ then models self-referencing in that it requires  all
elements of $B$ to compute any one element of $B^{-1}$. As well there is the 
additive SRN  
$w_{ij}$ which limits the self-referential information  but, significantly, also acts in such
a way that the network is innovative in the sense of generating semantic information, that is
information which is internally meaningful.  The emergent behaviour is believed to be
completely generic in that it is not suggested that reality is a computation, rather it
appears that reality has the form of an self-referential order-disorder
information system.  It is important to note that process physics is a non-reductionist
modelling of reality; the basic iterator (1) is premised on the general assumption  that
reality is sufficiently complex that self-referencing occurs, and that this has limitations.
Eqn.(1) is then a minimal bootstrapping implementation of these notions.  At higher emergent levels
this self-referencing manifests itself  as {\it interactions} between emergent patterns.    

To be a successful contender for the Theory of Everything (TOE) process
physics must ultimately prove the uniqueness conjecture:  that the  characteristics (but not
the contingent details) of the  pure  semantic information system are unique.  This would
involve demonstrating both the effectiveness of the SOC filter and the robustness of the
emergent phenomenology, and the complete agreement of the later with observation.    

The stochastic neural network is  modelled by the iterative process
\begin{equation}
B_{ij} \rightarrow B_{ij} -\alpha (B + B^{-1})_{ij} + w_{ij},  \mbox{\ \ } i,j=1,2,...,2M;
M
\rightarrow
\infty,
\label{eq:map}\end{equation}
 where 
 $w_{ij}=-w_{ji}$ are
independent random variables for each $ij$ pair and for each iteration and chosen from some probability
distribution. Here $\alpha$ is a parameter the precise value of which should not be critical but which
influences the self-organisational process. 
We start the iterator at 
$B\approx 0$, representing the absence of information.  
  With the noise absent the iterator 
behaves in a deterministic and reversible manner given by
the matrix
\begin{equation}
B =  MDM^{-1}; \mbox{\ \ \ \  }D=\left(\begin{array}{rrrrrr}
0 & +b_1 & 0 & 0\\
-b_1 & 0 & 0 & 0 \\
0 & 0 & 0 & +b_2 \\
0 & 0 & -b_2 & 0 & \\
 &&&&. \\ &&&&&. \\          
\end{array}\right),  \mbox{\ \ \ \ \  }  b_1,b_2,... \geq 0,
\end{equation}
where $M$ is  a real orthogonal matrix determined uniquely by the start-up $B$, and each $b_i$
evolves according to the iterator
$
b_i \rightarrow b_i-\alpha(b_i-b_i^{-1}),
$
which converges to $b_i=1$.  This $B$ exhibits no interesting structure. In the presence of the
noise the iterator process  is non-reversible and non-deterministic.  It   is also manifestly
non-geometric and non-quantum, and so does not assume any of the standard features of syntax based
physics models.   The dominant  mode is  the  formation of an apparently randomised 
background (in
$B_{ij}$) but, however, it  also manifests  a   self-organising process which results
in a growing  three-dimensional  fractal process-space that competes with this random background -
the formation of a `bootstrapped universe'.  Here we report on the current status of ongoing
work to extract the nature of this `universe'. 

The emergence of order in this system might
appear to violate expectations regarding the 2nd Law of Thermodynamics; however because of the SRN
the system behaves as an open system and the  growth of order arises from the
self-referencing term, $B^{-1}$ in (1), selecting certain implicit order in the SRN.  Hence the SRN
acts as a source of negentropy\footnote{The term {\it negentropy} was introduced by E.
Schr\"{o}dinger\cite{negentropy} in 1945, and since then there has been ongoing discussion of its 
meaning. In process physics it manifests as the SRN.}

 This  growing  three-dimensional  fractal process-space is an example of a Prigogine far-from-
equilibrium dissipative structure\cite{Prigogine} driven by the SRN. 
 From each iteration the noise term
will additively introduce rare large value
$w_{ij}$.  These  $w_{ij}$, which define sets of strongly linked nodes, will persist   through more
iterations than smaller valued $w_{ij}$ and, as well,  they become further linked  by the iterator
to form a three-dimensional process-space with embedded topological defects. In this way the
stochastic neural-network creates stable strange attractors and as well determines their interaction
properties. This information is all internal to the system; it is the semantic information within
the network.
 
 To see the nature of this internally generated information  consider a node $i$ involved in one
such large
$w_{ij}$;   
 it will be   connected via  other large $w_{ik}$ to a
number of other nodes and so on, and this whole set of connected nodes forms a connected random graph unit
which we call a gebit as it acts as a small piece or bit of geometry formed from random information links and 
from which the process-space is self-assembled. The gebits compete for new links  and undergo
mutations. Indeed, as will become clear, process physics is remarkably analogous in its operation to
biological systems. The reason for this is becoming clear: both reality and subsystems of reality must use
semantic information processing to maintain existence, and symbol manipulating systems are totally unsuited
to this need, and in fact totally contrived.

To analyse the connectivity of such 
gebits assume for simplicity that the large $w_{ij}$  arise with fixed but very small probability $p$,
then the geometry of  the gebits is revealed by studying the probability distribution for  the structure
of the random graph units or gebits minimal spanning trees with $D_k$ nodes  at $k$ links from node $i$
($D_0
\equiv 1$), see Fig.4c, this is given by\cite{Nagels}

\begin{equation}{\cal P}[D,L,N] \propto \frac{p^{D_1}}{D_1!D_2!....D_L!}\prod_{i=1}^{L-1}
(q^{\sum_{j=0}^{i-1}{D_j}})^{D_{i+1}}(1-q^{D_i})^{D_{i+1}},\end{equation}
where $q=1-p$, $N$ is the total number of nodes in the gebit and $L$ is the maximum depth from node $i$. 
To find the most likely connection pattern we numerically maximise ${\cal P}[D,L,N]$ for fixed $N$ with respect
to
$L$ and the $D_k$. The resulting $L$ and $\{D_1,D_2,...,D_L\}$ fit very closely to the form $D_k\propto
\sin^{d-1}(\pi k/L)$;  see Fig.5a  for $N=5000$ and $\mbox{Log}_{10}p=-6$.  The resultant  $d$
values for a range of $\mbox{Log}_{10}p$ and $N=5000$ are shown in Fig.5b. 

\vspace{-35mm}
\hspace{-20mm}\includegraphics{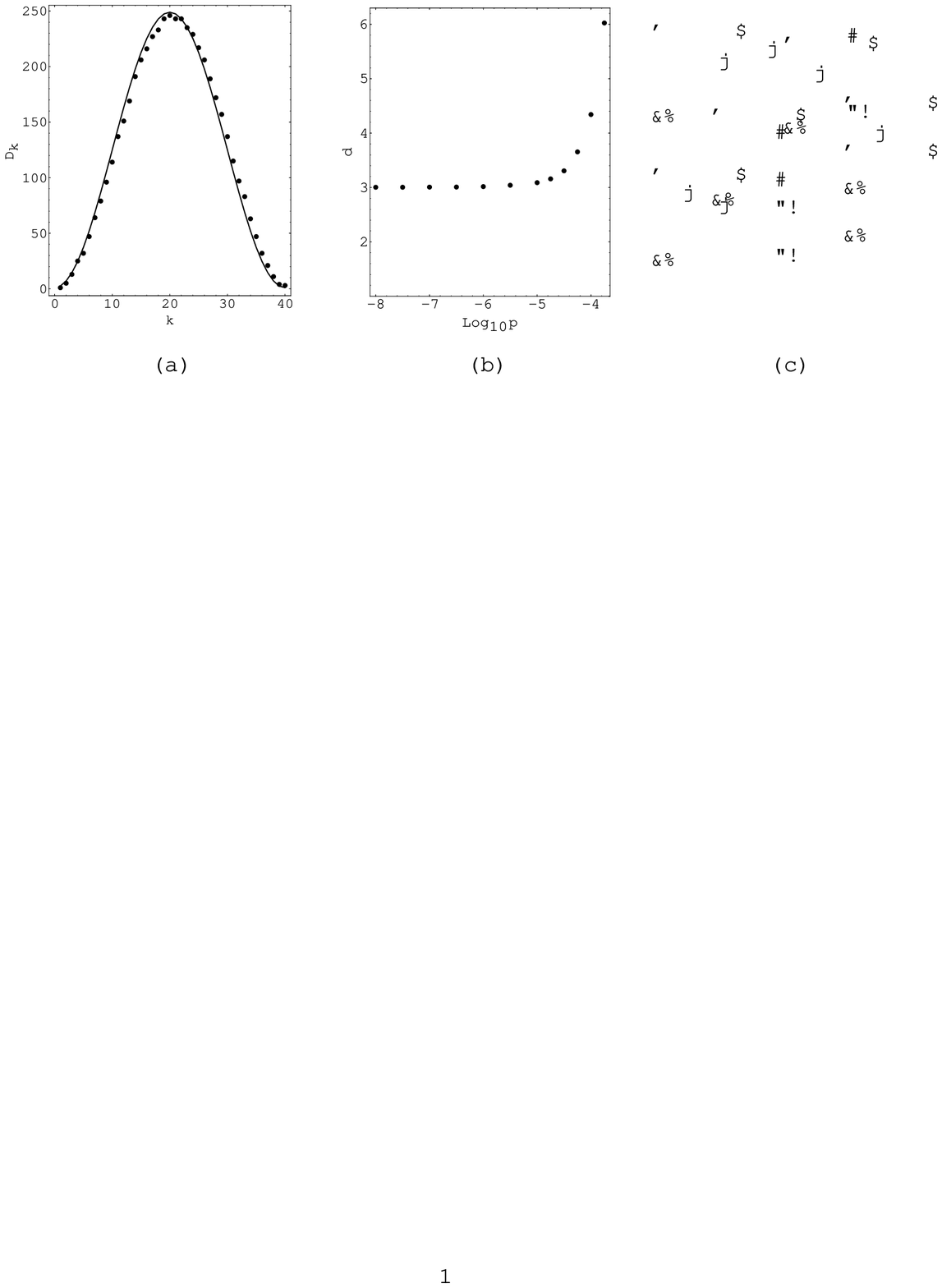}

\vspace{70mm}
\begin{figure}[h]
\caption{\small 
(a) Points show the $D_k$ set and $L=40$ value found by numerically  maximising ${\cal P}[D,L,N]$
for $\mbox{Log}_{10}p=-6$ for fixed  $N=5000$. Curve shows
$D_k\propto \sin^{d-1}(\frac{\pi k}{L})$ with best fit $d=3.16$ and $L=40$, showing  excellent
agreement, and indicating  embeddability in an $S^3$ with some topological defects. (b) Dimensionality $d$ of
the gebits as a function of  the probability $p$. (c) Graphical depiction of the `process space' at one stage of the iterative
process-time showing a quantum-foam structure formed from  embeddings and linkings of gebits.  The
linkage connections have the distribution of a 3D space, but the individual gebit components are
closed compact spaces and cannot be embedded in a 3D background space.  So the drawing is only
suggestive. Nevertheless this figure indicates that process physics generates a cellular
information system, where the behaviour is determined at all levels by internal information. } 
\end{figure}

This shows, for $p$ below a critical value, that
$d=3$, indicating that the connected nodes have a natural embedding in a 3D hypersphere $S^3$;
call this a base gebit. Above that value of $p$,   the increasing value of $d$
indicates the presence of extra links that, while some conform with the embeddability,  are in the main defects
with respect to the geometry of the $S^3$.  These extra links act as topological defects.  By themselves these
extra links will have the   connectivity and embedding geometry of numbers of gebits, but these gebits have a
`fuzzy' embedding in the base gebit. This is an indication of  fuzzy  homotopies  (a homotopy is,
put simply, an embedding of one space into another).  

The base gebits $g_1, g_2, ...$ arising from the SRN together with their embedded topological defects have
another remarkable property:  they are `sticky' with respect to the iterator.  Consider the   larger valued 
$B_{ij}$ within a given gebit  $g$, they form  tree graphs and most tree-graph adjacency matrices are singular
 (det($g_{tree})=0$).  However  the presence of other smaller valued $B_{ij}$ and the
general
 background noise  ensures that det$(g)$ is small  but not exactly zero. 
Then the 
$B$ matrix has an inverse with large components that act to cross-link  the new and
existing gebits. This cross-linking is itself random, due to the presence of background noise, and the above
analysis  may again be used and we would conclude that  the base gebits themselves are formed into a 3D hypersphere
with embedded topological defects.  The nature of the resulting 3D process-space is suggestively indicated in Fig.5c, and
behaves essentially as a quantum foam\cite{foam1}.

Over ongoing
iterations the existing gebits become cross-linked and eventually lose their ability to undergo further
linking; they lose their `stickiness'  and decay. The value of the parameter $\alpha$ in (1) must
be small enough that the `stickiness' persists over many iterations, that is, it is not quenched
too quickly, otherwise the emergent network  will not grow.  Hence the emergent
space is 3D but is continually undergoing  replacement of its component gebits;  it is an
informational process-space, in sharp distinction to the non-process continuum geometrical spaces
that have played a dominant role in modelling physical space.  If the noise is `turned off' then
this emergent dissipative space will decay and cease to exist.  We thus see that the nature of
space is deeply related to the logic of the limitations of logic, as implemented here as a
self-referentially limited neural network.

\section{Modelling Gebits and their Topological Defects}

We need to extract convenient but approximate syntactical descriptions  of the semantic information in the
network, and these will have the form of a sequence of  mathematical constructions, the first being the Quantum
Homotopic Field Theory. Importantly they must all retain explicit manifestations of the SRN. To this  end first
consider the special case of the iterator when the SRN is frozen at a particular
$w$, that is we consider iterations with  an artificially  fixed SRN term. Then the iterator is equivalent to the
minimisation of an `energy' expression (remember that
$B$ and $w$ are antisymmetric)
\begin{equation}
E[B;w]= -\frac{\alpha}{2}\mbox{Tr}[B^2]-\alpha \mbox{TrLn}[B]+\mbox{Tr}[wB].
\end{equation}
Note that for disconnected gebits $g_1$ and $g_2$ this energy is additive, $E[g_1\oplus g_2]=E[g_1]+E[g_2]$.
Now suppose the fixed $w$ has the form of a  gebit approximating an $S^3$ network with one embedded topological
defect which is itself an $S^3$ network, for simplicity.  So we are dissecting the gebit into base gebit, defect gebit
and linkings or embeddings between the two. We also ignore the rest of the network, which is permissible if our gebit
is disconnected from it.  Now  if det$(w)$ is not small, then this gebit is non-sticky, and for small $\alpha$, the
iterator converges to $B\approx\frac{1}{\alpha}w$, namely an enhancement only of the gebit.  However because the
gebits are rare constructs they tend to be composed of larger $w_{ij}$ forming tree structures, linked by smaller
valued
$w_{ij}$.  The tree components make det$(w)$ smaller, and then the inverse $B^{-1}$ is activated and generates new
links.  Hence, in particular, the topological defect relaxes, according to
the `energy' measure, with respect to the base gebit.  This relaxation  is an example of a `non-linear elastic'
process\cite{Ogden}.  The above gebit has the form of  a mapping  $\pi: S \rightarrow \Sigma$ from a base space to a
target space.  Manton\cite{Manton1, Manton2, GP} has constructed the continuum form for the `elastic energy' of such
an embedding and for $\pi: S^3 \rightarrow S^3$  it is the Skyrme energy
\begin{equation}
E[U]=\int \left[ -\frac{1}{2}\mbox{Tr}(\partial_i UU^{-1}\partial_i UU^{-1}) -\frac{1}{16} \mbox{Tr}[\partial_i
UU^{-1},\partial_i UU^{-1}]^2\right],
\end{equation}
where $U(x)$ is an element of $SU(2)$. Via the parametrisation $U(x)=\sigma(x)
+i\vec{\pi}(x).\vec{\tau}$, where the
$\tau_i$ are Pauli matrices,  we have $\sigma(x)^2+\vec{\pi}(x)^2$=1, which parametrises an $S^3$ as
a unit hypersphere  embedded in $E^4$. Non-trivial minima of
$E[U]$ are known as Skyrmions (a form of topological soliton), and have
$Z=\pm1,\pm2,...$, where $Z$ is the winding number of the map, 
\begin{equation}
Z=\frac{1}{24\pi^2}\int\sum\epsilon_{ijk}\mbox{Tr}(\partial_i UU^{-1}\partial_j UU^{-1}\partial_k UU^{-1}).
\end{equation}

The first key to extracting  emergent phenomena from the stochastic neural network is the validity of this continuum
analogue, namely that $E[B;w]$ and $E[U]$ are describing essentially the same `energy'
reduction process.  This should be amenable to detailed  analysis.  

This `frozen' SRN analysis of course does not match the time-evolution of the
full iterator, (1), for  this displays a much richer collection of processes.  With ongoing new noise in each
iteration and the saturation of the linkage possibilities of the gebits emerging from this noise, there arises a
process of ongoing birth, linkaging and then decay of most patterns.  The task is then to
identify those particular patterns that survive this flux, even though all components of
these patterns eventually disappear, and to attempt a description of their modes of
behaviour.  This brings out the very biological nature of the information processing in
the SNN, and which appears to be characteristic of a `pure' semantic information system.

In general each gebit, as it emerges from the SRN, has active nodes and embedded topological defects, again with
active nodes.  Further there will be defects embedded in the defects and so on,  and so gebits begin to have the
appearance of a fractal defect structure, and all the defects having various classifications and associated  winding
numbers.  The energy analogy above suggests that  defects with opposite  winding numbers at the same fractal depth 
may annihilate by drifting together and merging. Furthermore the embedding of the defects is unlikely to be
`classical', in the sense of being described by a mapping
$\pi(x)$, but rather would be fuzzy, i.e described by some functional, $F[\pi]$, which would correspond to a
classical embedding only if $F$ has a supremum at one particular $\pi=\pi_{cl}$. As well these gebits are undergoing
linking because their active nodes (see \cite{CK98} for more discussion) activate the $B^{-1}$ new-links
process between them, and so by analogy the gebits themselves form larger structures with embedded fuzzy
topological defects. This emergent behaviour is suggestive of a quantum  space foam, but one containing
topological defects which will be preserved by the system, unless annihilation events occur.  If these
topological defects are sufficiently rich in fractal structure as to be preserved, then their initial
formation would have occurred as the process-space relaxed out of its initial, essentially random form.
This phase would correspond to the early stages of the Big-Bang. Once the topological defects are trapped
in the process-space  they are doomed to meander through that space by essentially self-replicating, i.e.
continually having their components die away and be replaced by similar components.  These residual
topological defects are what we call matter. The behaviour of both the process-space and its defects is
clearly determined by the same network processes; we have an essential unification of space and matter
phenomena.  This emergent quantum foam-like behaviour suggests that the full generic  description of the
network behaviour  is via the Quantum Homotopic Field Theory (QHFT) of the next section.

\section{Modelling the Emergent Quantum Foam}

To construct this QHFT we introduce an appropriate configuration space, namely  all the possible homotopic 
mappings $\pi_{\alpha\beta}: S_\beta \rightarrow S_\alpha$, where the 
$S_1,S_2,..$, describing `clean' or topological-defect free gebits, are compact spaces of various types. Then  QHFT
has the form of an iterative functional Schr\"{o}dinger equation for the discrete time-evolution of a wave-functional 
$\Psi[....,\pi_{\alpha\beta},....;t]$
\begin{equation}\Psi[....,\pi_{\alpha\beta},....;t+\Delta
t]= \Psi[....,\pi_{\alpha\beta},....;t]
-iH\Psi[....,\pi_{\alpha\beta},....;t]\Delta t +  \mbox{QSD terms},
\end{equation}       The time step $\Delta t$ in (7)  is relative to the scale of the fractal
processes being explicitly described, as we are using a configuration space of mappings between prescribed gebits.  At
smaller scales we would need a smaller value of   $\Delta t$.  Clearly this invokes a (finite)
renormalisation scheme. We now discuss the form of the hamiltonian and the Quantum State Diffusion (QSD) terms.

First (7), without the QSD term,  has a form analogous to  a `third quantised' system, in
conventional terminology\cite{Baby}. These 
systems were considered   as   perhaps capable of generating  a quantum theory of
gravity. The argument here  is that this is the emergent behaviour of the SNN, and it does indeed lead to quantum
gravity, but with quantum matter as well.  More importantly we understand the origin of (7), and it will lead to
quantum and then classical gravity, rather than arise from classical gravity via some ad hoc or heuristic
quantisation procedure.

Depending on the `peaks' of 
$\Psi$ and the connectivity of the resultant dominant mappings such mappings are to be interpreted as
either embeddings  or links; Fig.5c then suggests the dominant process-space form within
$\Psi$ showing both links and embeddings. The emergent process-space then has the characteristics of a  
quantum foam. Note that, as indicated in Fig.5c, the original start-up links and nodes are now absent.
Contrary to the suggestion in Fig.5c, this process space cannot be embedded  in a {\it finite}
dimensional geometric space with the emergent metric preserved, as it is composed of  nested
or fractal finite-dimensional closed spaces.  

We now consider the form of the hamiltonian H.  The previous section suggested that Manton's non-linear
elasticity interpretation of the Skyrme energy is appropriate to the SNN.  This then suggests that H is
the functional operator
\begin{equation}
H=\sum_{\alpha\neq\beta}h[\frac{\delta}{\delta
\pi_{\alpha\beta}},\pi_{\alpha\beta}]
\end{equation}
where $h[\frac{\delta}{\delta \pi},\pi]$ is the (quantum) Skyrme Hamiltonian functional operator for the system based
on  making fuzzy the  mappings
$\pi: S \rightarrow \Sigma$, by having $h$ act on wave-functionals of the form $\Psi[\pi(x);t]$. Then $H$
is the sum of pairwise  embedding or homotopy hamiltonians. The corresponding functional Schr\"{o}dinger
equation would simply  describe the time evolution of quantised Skyrmions with the base space fixed, and
$\Sigma \in SU(2)$. There have been very few analyses of even this class of problem, and then the base space
is usually taken to be $E^3$.  We shall not give the explicit form of $h$ as it is complicated, but wait to present
the associated action. 

In the absence of the QSD terms the time evolution in (7) can be formally written as a functional integral
\begin{equation}
\Psi[\{\pi_{\alpha\beta}\};t']=\int\prod_{\alpha\neq\beta}{\cal
D}\tilde{\pi}_{\alpha\beta}e^{iS[\{\tilde{\pi}_{\alpha\beta}\}]}
\Psi[\{\pi_{\alpha\beta}\};t],
\end{equation}
where, using the continuum $t$ limit notation, the action is a sum of pairwise actions,
\begin{equation}
S[\{\tilde{\pi}_{\alpha\beta}\}]=\sum_{\alpha\neq\beta}S_{\alpha\beta}[\tilde{\pi}_{\alpha\beta}],
\end{equation}
\begin{equation}
S_{\alpha\beta}[\tilde{\pi}]=\int_t^{t'}dt''\int d^nx\sqrt{ -g} \left[ \frac{1}{2}\mbox{Tr}(\partial_\mu
\tilde{U}\tilde{U}^{-1}\partial^\mu
\tilde{U}\tilde{U}^{-1}) +\frac{1}{16} \mbox{Tr}[\partial_\mu \tilde{U}\tilde{U}^{-1},\partial^\nu
\tilde{U}\tilde{U}^{-1}]^2\right],
\end{equation}
and the now time-dependent (indicated by the tilde symbol) mappings $\tilde{\pi}$ are parametrised by
$\tilde{U}(x,t)$, $\tilde{U}\in S_\alpha$. The metric $g_{\mu\nu}$ is that of the $n$-dimensional base space,
$S_\beta$, in
$\pi_{\alpha,\beta}: S_\beta
\rightarrow  S_\alpha$. As usual in the functional integral formalism the functional derivatives in the quantum
hamiltonian, in (8), now manifest as the time components $\partial_0$ in (11),  so now (11) has the form
of a `classical' action, and we see the emergence of `classical' fields, though the emergence of `classical' behaviour
is a more complex  process.  Eqns.(7) or (9) describe an infinite set of quantum skyrme systems, coupled in a 
pairwise manner.  Note that each homotopic mapping appears in both orders; namely $\pi_{\alpha\beta}$ and 
$\pi_{\beta\alpha}$.

The   Quantum State Diffusion (QSD)\cite{QSD} terms  are non-linear and
stochastic, 
\begin{equation}
\mbox{QSD terms} =\sum_\gamma\left(
<\!\!L^\dagger_\gamma\!\!>L_\gamma-\frac{1}{2}L^\dagger_\gamma L_\gamma-
<\!\!L^\dagger_\gamma\!\!><\!\!L_\gamma\!\!>\right)
\Psi\Delta t+\sum_\gamma\left(L_\gamma-<\!\!L_\gamma\!\!> \right)\Psi\Delta \xi_\gamma,
\end{equation}
which involves summation over the class of Linblad functional operators $L_\gamma$.
The QSD terms  are  up to 5th order in $\Psi$, as in general,
\begin{equation}
<\!\! A \!\!>_t=\int \prod_{\alpha\neq\beta}{\cal D}\pi_{\alpha\beta}\Psi[\{
\pi_{\alpha\beta} \};t]^* A 
\Psi[\{ \pi_{\alpha\beta}\};t]
\end{equation}
and where $\Delta \xi_\gamma$
are complex statistical variables with means  $M (\Delta \xi_\gamma) = 0$, $M( \Delta
\xi_\gamma
\Delta\xi_{\gamma'})= 0$  and $M(\Delta \xi^*_\gamma\Delta\xi_{\gamma'}) =
\delta(\gamma-\gamma')\Delta t$

These QSD terms are ultimately responsible for the emergence of
classicality via an  objectification process, but in particular  they produce
wave-function(al) collapses during quantum measurements, as the QSD terms tend to `sharpen' the fuzzy homotopies
towards  classical or sharp  homotopies (the forms of the Linblads will be discussed in detail elsewhere).   So the
QSD terms, as residual SRN effects, lead to 
 the Born quantum measurement  random behaviour, but here arising from the process physics, and not being
invoked as a metarule.
 Keeping the QSD terms leads to a functional integral representation for a
density matrix formalism in place of (9), and this amounts to a derivation of  the decoherence
formalism which is usually arrived at by invoking the Born measurement metarule. Here we see that
`decoherence'  arises from the limitations on self-referencing.

In the above we have  a deterministic and unitary
evolution, tracking and preserving topologically encoded information, together with the stochastic
QSD terms, whose form protects that information during localisation events, and which
also  ensures the full matching in QHFT of process-time to real time: an ordering of
events, an intrinsic direction or `arrow' of time and a modelling of the contingent
present moment effect.   So we see that process physics generates a complete theory of quantum
measurements involving the  non-local, non-linear and  stochastic QSD terms.  It does this because
it generates both the `objectification' process associated with the classical apparatus and the actual
process of (partial)  wavefunctional collapse as the quantum modes interact with the measuring
apparatus.  Indeed many of the mysteries of quantum measurement are resolved when it is realised that
it is the measuring apparatus itself that actively provokes the collapse, and it does so because the
QSD process is most active when the system deviates strongly from its dominant mode, namely the
ongoing relaxation of the system to a 3D process-space.  This is essentially the process that
Penrose\cite{Penrose} suggested, namely that the quantum measurement process is essentially a
manifestation of quantum gravity. The demonstration of the validity of the Penrose argument of course
could only come about when  quantum gravity was {\it derived} from deeper considerations, and not by
some {\it ad hoc} argument such as the {\it quantisation} of Einstein's classical spacetime model.

The mappings  $\pi_{\alpha\beta}$ are related to group manifold parameter spaces with the group determined by
the dynamical stability of the mappings. This symmetry leads to  the flavour symmetry of the
standard model. Quantum homotopic mappings or skyrmions  behave as fermionic or bosonic modes for 
appropriate winding numbers; so process physics predicts both fermionic and bosonic quantum modes, but
with these  associated with topologically encoded information and not with  objects or `particles'. 

\section{Quantum Field Theory}
The QHFT is a very complex `book-keeping' system for the emergent properties of the neural network,
and we now sketch how we may extract a more familiar quantum field theory (QFT) that relates to the
standard model of `particle' physics. An effective QHFT should reproduce the emergence
of the process-space part of the quantum foam, particularly its 3D aspects. The QSD processes play
a key role in this as they tend to enhance classicality. Hence at an appropriate scale QHFT
should approximate to a more conventional QFT, namely the emergence of a wave-functional system
$\Psi[U(x);t]$ where the  configuration space is that of homotopies from a 3-space to $U(x) \in
G$, where $G$ is some group manifold space. This $G$ describes `flavour' degrees of freedom. 
 Hence the Schr\"{o}dinger
wavefunctional equation for this QFT  will have the form
\begin{equation}\Psi[U;t+\Delta
t]= \Psi[U;t]
-iH\Psi[U;t]\Delta t +  \mbox{QSD terms},
\end{equation} 
where the general form of $H$ is known, and where a new residual manifestation of the SRN appears
as the QSD terms.  This system describes skyrmions embedded in a continuum spacetime. It is
significant that such Skyrmions are only stable, at least in flat space and for static skyrmions, if
 that space is 3D.  This tends to confirm the observation that 3D space is special for the neural
network process system.
Again, in the absence of the QSD terms, we may express (15) in terms of the functional integral
\begin{equation}
\Psi[U;t']=\int{\cal
 D}\tilde{U}e^{iS[\tilde{U}]}
\Psi[U;t].
\end{equation}
To gain some insight into the phenomena present in (14) or (15), it is convenient to use the
fact that functional integrals of this Skyrmionic form my be written in terms of  Grassmann-variable
functional integrals, but only by introducing a fictitious `metacolour' degree of freedom and 
associated coloured fictitious  vector bosons. This is essentially the reverse of the Functional Integral Calculus
(FIC) hadronisation technique in  the Global Colour Model (GCM) of QCD\cite{GCM}.  The action for the Grassmann and
vector boson  part of the system is of the form (written for flat spacetime)
\begin{equation}
S[\overline{p},p,A^a_\mu]=\int d^4x\left( \overline{p}\gamma^\mu(i\partial_\mu
+g\frac{\lambda^a}{2}A^a_\mu)p-\frac{1}{4}F^a_{\mu\nu}(A)F^{a\mu\nu}(A)
\right),
\end{equation}
where the  Grassmann variables $p_{f c}(x)$ and $\overline{p}_{f c}(x)$ have
flavour and metacolour labels. The Skyrmions are then the low energy  Nambu-Goldstone
modes of   this Grassmann system; other emergent modes are of higher energy and can be
ignored.  These coloured and flavoured but fictitious fermionic fields
$\overline{p}$ and $p$  correspond to the proposed preon system\cite{Preons1,Preons2}. As
they are purely fictitious, in the sense that there  are no excitations in the system corresponding
to them, the metacolour degree of freedom must be hidden or confined. Then while the QHFT and the QFT
represent an induced  syntax for the semantic information, the preons may be considered as an
induced `alphabet' for that syntax. The advantage of introducing this preon alphabet is that we can
more easily determine the states of the system by using  the more familiar  language of fermions and
bosons, rather than working with the skyrmionic system, so long as only colour singlet states
are finally permitted.  However it is important to note that (16) and the action in (15) are certainly not
the final forms.  Further analysis will be required to fully extract the induced
actions for the emergent QFT.  

\section{Inertia and Gravity}
Process physics predicts that the neural network behaviour will be charactersied by a growing
3-dimensional process-space having, at a large scale, the form of an $S^3$ hypersphere, which is
one of the forms allowed by Einstein's syntactical modelling. It is possible to give the dominant
rate of growth of this hypersphere. However first, from  random graph theory\cite{randomg}, we
expect more than one such spatial system, with 
each having the character of a growing hypersphere, and all embedded in the random background
discussed previously. This background has no metric structure, and so these various hyperspheres
have no distance measure over them. We have then a multi-world universe (our `universe' being
merely one of these `worlds'). Being process spaces they compete for new gebits, and so long as we
avoid a saturation case, each will grow according to
\begin{equation}
\frac{dN_i}{dt}=aN_i-bN_i  \mbox{\ \ \ \ \ } a>0, b>0,
\end{equation}
where the 
last term describes the decay of gebits at a rate $b$, while the first describes growth of the
$i$-th `world', this  being proportional to the size  (as measured by its gebit content  number)
$N_i(t)$ as success in randomly attaching new gebits is proportional to the number of gebits 
present (the `stickiness' effect), so long as we ignore the topological defects (quantum `matter')
as these have a different stickiness, and also affect the decay rate, and so slow down the
expansion. Thus
$N_i(t)$ will show exponential growth, as appears to be the case as indicated by recent observations
of very distant supernovae counts\cite{cosmo}. Hence process physics predicts a 
positive  cosmological constant. 

One striking outcome of process physics is a possible explanation for the phenomenon of gravity. 
First note that matter is merely topological defects embedded in the process space, and we expect
such defects to have a larger than usual gebit connectivity; indeed matter is a violation of the
3-D connectivity of this space, and it is for this reason we needed to introduce fields to emulate
this extra non-spatial connectivity.  One consequence of this is that in the region of these
matter fields the gebits decay faster, they are less sticky because of the extra connectivity. 
Hence in this region, compared to other nearby matter-free regions the gebits are being `turned
over' more frequently  but at the same time are less effective in attracting new gebits. Overall
this suggests that  matter-occupying-regions act as net sinks for gebits, and there will be a
trend for the neighbouring process-space to undergo a diffusion/relaxation process in which
the space effectively moves towards the matter: matter acts as a sink for space, and never as a
source. Such a process would clearly correspond to gravity. As  the effect is described by
a net diffusion/relaxation of space which  acts equally on all surrounding matter, the
in-fall mechanism is independent of the nature of the surrounding matter. This is nothing
more than Einstein's Equivalence Principle.   As well if the in-fall rate exceeds the rate at which
`motion' through the process-space is possible then an event horizon appears, and this  is clearly  the
black hole scenario. Such an event horizon is sufficient condition for the occurrence of  Hawking
radiation.

Finally we mention one long standing unsolved problem in physics, namely an understanding of
inertia. This is the effect where objects continue in uniform motion  unless acted upon by `forces',
and was first analysed by Galileo. However there has never been an explanation for this effect;
in Newtonian physics it was built into the syntactical description rather than being a prediction
of that modelling.  Of course current physics is essentially a static modelling of reality, with
motion indirectly accessed via the geometrical-time metarule, and so the failure to explain motion
is not unexpected. However process physics offers a simple explanation.

The argument for inertia follows from a simple self-consistency argument. Suppose a 
topological defect, or indeed a vast bound collection of such defects, is indeed `in motion'. This
implies that the gebits are being preferentially replaced  in the direction of that `motion', for
motion as such is a self-replication process; there is no mechanism in process physics for a fixed
pattern to say `slide' {\it through} the process-space.  Rather motion is self-replication of the
gebit  connectivity patterns in a set direction.  Since the newest gebits, and hence the stickiest
gebits, in each topological defect,  are on the side corresponding to  the direction of motion, the
 gebits on that side are preferentially replaced.  Hence the motion is self-consistent and
self-perpetuating.

An additional effect expected in process physics is that such motion results in a time dilation
 effect;  the self-replication effect  is to be considered as partly  the self-replication  associated
with any internal oscillations and partly self-replication associated with `motion'. This `competition for
resources' results in the slowing down of internal oscillations, an idea discussed by 
Toffoli\cite{Toffoli}. We then see that many of the effects essentially assumed by the formalism of general
relativity appear to be emergent phenomena within process physics. 

However there is one novel effect that is of some significance. The  process-space appears to
represent a preferred frame of reference, but one that is well hidden by the time-dilation effects.
Hence we would not expect a Michelson-Morley type experiment  to reveal this frame.  However using
the arguments of  Hardy\cite{Hardy}  we  expect that
the action of the QSD wavefunctional `collapse' processes would reveal the proper frame 
through a multi-component   EPR experiment, as the non-local QSD collapse occurs in a truly
simultaneous manner; essentially it exposes the underlying global iterations.

\section{Conclusions}
Sketched out here is the radical proposal that to comprehend reality we need a system richer than
mere  syntax to capture the notion that reality is at all levels  about, what may be
called,  internal, relational or semantic information, and not, as in the case of syntax,
information that is essentially accessible to or characterisable by  observers.   This 
necessitates an evolution in modelling reality from a non-process  physics to a process
physics.  Such a development has been long anticipated outside of physics  where it is
known as {\it Process Philosophy}\cite{ProcessPhilosophy}.

To
realise such a system one representation has been proposed and studied, namely that of a
self-referentially limited neural-network model, for neural networks are powerful examples of
non-symbolic information processing. This neural-network model is manifestly free of any notions
of geometry, quantum phenomena or even `laws of physics'. Nevertheless the arguments presented here
strongly suggest that such phenomena are emergent in the neural network but only because the
self-referential noise acts as a source of  negentropy or order.  The model has been developed to the
stage where various phenomena have been identified and appropriate induced syntactical descriptions
have been suggested. These correspond essentially to the concepts of current physics which, over
the years, have been arrived at via increasingly more abstract non-process syntactical modelling.  One
important addition being the ever-present QSD terms which, as it happens, ensure that the phenomena of
time fully matches our experiences of time, and which also plays various other key roles.  This new
process-physics is inherently non-reductionist as it explicitly assumes that reality is sufficiently
complex that it is self-referential, and which may be accessed by using a subtle  bootstrap approach. 
Clearly we see the beginnings of a unification of physics that leads to quantum gravity and classicality
and the emergence of syntax and its associated logic of named objects.  The confirmation of process
physics will involve developing more powerful techniques to facilitate the extraction of the induced
syntax for the emergent phenomena. This is a novel and subtle problem. 

Process physics also implies that the geometrical construct of spacetime is merely an induced formal system or syntax
and does not itself have any ontological status. Nevertheless this syntax is enormously useful for
analysing certain technical aspects of reality, so long as we can be sure that the syntax alone does not
introduce spurious problems. One well known spurious problem are the divergences in quantum field theory
caused by the continuum nature of spacetime. As well it is obvious that we can never discover deeper
physics by {\it quantising} classical systems. 

The author thanks the organisers of this conference for assembling an interesting collection of
contributions, and in particular expresses his regards to Dr Peter Szekeres on his retirement
after a distinguished career in general relativity studies.

\end{document}